\documentclass[twoside,reqno]{bjp}
\usepackage{graphics}
\usepackage[pdftex]{graphicx}
\usepackage{amssymb,amsmath}
\usepackage{times}
\newcommand\be{\begin{equation}}
\newcommand\ee{\end{equation}}
\newcommand{\bea}{\begin{eqnarray}}
\newcommand{\eea}{\end{eqnarray}}

\newcommand{\nn}{\nonumber}
\newcommand{\pd}{\partial}

\begin{document}

\sloppy \raggedbottom

 \setcounter{page}{1}


% use the thanks command within \title, \author or \address for footnotes;
% \title{Title} or  \title{Title\thanks{...}}
% \author{Name1}{aff.label1}, \coauthor{Name2}{aff.label2},  \coauthor{Name3}{aff.label3}
% \address{Address1}{aff.label1}
% \address{Address2}{aff.label2}
% \address{Address3}{aff.label3}
%\runningheads{NAMES OF AUTHORS IN CAPITALS}{SHORT TITLE IN CAPITALS}

\title{Toward a Gravity Dual of Glueball Inflation\thanks{Invited talk at the memorial symposium for Prof. Matey Mateev, April 2015, Sofia, Bulgaria. L.A. is very grateful to Prof. Mateev for his insightful lectures in Sofia University.}}

\runningheads{$dS_4$ in Gauge/Gravity Duality}{L.~Anguelova,
P.~Suranyi, L.C.R.~Wijewardhana}

\begin{start}
\author{L.~Anguelova}{1}, \coauthor{P.~Suranyi}{2},
\coauthor{L.~C.~R.~Wijewardhana}{2}

\address{Institute of Nuclear Research and Nuclear Energy,
BAS, Sofia 1784, Bulgaria}{1}

\address{Department of Physics, University of Cincinnati, OH, 45221, USA}{2}

%\received{}

\begin{Abstract}
We summarize and extend our work on nonsupersymmetric solutions of a 5d consistent truncation of type IIB supergravity, that is relevant for gauge/gravity duality. The fields in this 5d theory are the five-dimensional metric and a set of scalars. We find solutions of the 5d equations of motion, which represent $dS_4$ foliations over the fifth (radial) dimension. In each solution at least one scalar has a nontrivial radial profile. These scalars are interpreted as glueballs in the dual gauge theory, living in 4d de Sitter space. We explain why this lays a foundation for building gravity duals of glueball inflation models.
\end{Abstract}

%11.25.Tq - Gauge/string duality; 98.80.Cq - Early Universe, Inflationary Universe, cosmic strings, domain walls in cosmology, particle theory models (Early Universe); 98.80.Bp - Big Bang theory, origin and formation of the Universe
\end{start}

\section[]{Introduction}

Gauge/Gravity Duality is a new powerful method for the study of strongly coupled gauge theories. It originated with the Maldacena conjecture, widely known as AdS/CFT correspondence \cite{JM}. Since then it has been extended to vast classes of theories, which are not conformal and whose gravitational duals are not given by $AdS$ space. Most of this work, however, applies only to supersymmetric field theories living in Minkowski space. Much less is known about strongly-coupled gauge theories in curved backgrounds. In \cite{ASW} we investigated a gravity dual for field theories in 4d de Sitter space.

The starting point of this investigation is the five-dimensional consistent truncation of type IIB supergravity, established in \cite{BHM}. The gravity duals of ${\cal N} = 1$ SYM, provided by the famous Maldacena-Nunez \cite{MN} and Klebanov-Strassler \cite{KS} solutions, as well as their more recent deformations \cite{NPP,ENP,EGNP}, can all be found as special solutions of the equations of motion of this five-dimensional theory. However, we are interested in solutions which are non-supersymmetric and represent $dS_4$ fibrations over the fifth (radial) direction. 

In addition to providing a playground for studying strongly-coupled gauge theories in de Sitter space, such solutions have great importance for models of Cosmological  Inflation, in which the inflaton is a composite state. One reason is that, of course, de Sitter space is the leading approximation to spacetime during Inflation. But, more importantly, we will argue here that, by considering an expansion in small time-dependent deformations around the above $dS_4$ foliations over the radial direction, one could build gravity duals of glueball inflation.

\section{The consistent truncation}

Let us begin by recalling that a consistent truncation of a $D$-dimensional theory to a $d<D$ dimensional one means by definition that {\it every} solution of the lower-dimensional theory lifts to a solution of the higher-dimensonal one. And the lift is entirely determined by the ansatz for the consistent truncation. Therefore, having a 5d consistent truncation of type IIB supergravity simplifies significantly the search for solutions of the latter, albeit of a certain kind. 

The 5d consistent truncation of type IIB, that is of interest here, is the one established in \cite{BHM}. It is given by the following ansatz of the 10d bosonic fields $g_{AB}$, $\phi$, $C$, $H_3$, $F_3$ and $F_5$ in terms of 5d ones:
\bea \label{10dmetric}
ds_{10d}^2 &=& e^{x+g} (\omega_1^2 \,+ \,\omega_2^2) \,+ \,e^{x-g} \!\left[ (\tilde{\omega}_1 + a \omega_1)^2 \!+ \!(\tilde{\omega}_2 - a \omega_2)^2 \right] \\
&+& \, e^{-6p-x} (\tilde{\omega}_3 + \omega_3)^2 + e^{2p-x} ds_{5d}^2 \quad , \quad s_{5d}^2 = g_{IJ} \,dx^I dx^J \quad , \nn
\eea
where
\bea \label{omegatilde}
\tilde{\omega}_1 &=& \cos \psi d\tilde{\theta} + \sin \psi \sin \tilde{\theta} d \tilde{\varphi} \,\, , \hspace*{1.5cm} \omega_1 = d \theta \,\, , \nn \\
\tilde{\omega}_2 &=& - \sin \psi d\tilde{\theta} + \cos \psi \sin \tilde{\theta} d \tilde{\varphi} \,\, , \hspace*{1.2cm} \omega_2 = \sin \theta  d \varphi \,\, , \nn \\
\tilde{\omega}_3 &=& d \psi + \cos \tilde{\theta} d \tilde{\varphi} \,\, , \hspace*{3cm} \omega_3 = \cos \theta d \varphi \,\, ,
\eea
and the RR 3-form is:
\bea
F_3 &=& P \left[ - (\tilde{\omega}_1 + b \,\omega_1) \wedge (\tilde{\omega}_2 - b \,\omega_2) \wedge (\tilde{\omega}_3 + \omega_3) \right. \\ 
&+& \left. \,(\pd_I b) \,dx^I \!\wedge (- \omega_1 \wedge \tilde{\omega}_1 + \omega_2 \wedge \tilde{\omega}_2) + (1-b^2) (\omega_1 \wedge \omega_2 \wedge \tilde{\omega}_3) \right] , \nn
\eea
where $P=const$, while the remaining fields are:
\be
\phi = \phi (x^I) \qquad , \qquad C = 0 \qquad , \qquad H_3 = 0
\ee
and
\be \label{F5}
F_5 = {\cal F}_5 + \star {\cal F}_5 \qquad , \qquad {\cal F}_5 = Q \,vol_{5d} \qquad , \qquad Q=const \quad .
\ee
In (\ref{10dmetric})-(\ref{F5}) the quantities $p$, $x$, $g$, $a$, $b$ and $\phi$ are scalars dependent on the 5d coordinates $x^I$. In principle, one can have a consistent truncation with $H_3 \neq 0$, as is the case in the Klebanov-Strassler solution \cite{KS}; see \cite{BHM}. But, for simplicity, we will work with the subcase of vanishing $H_3$ flux, which encompasses the Maldacena-Nunez solution \cite{MN} and its deformations \cite{NPP,ENP}. This is also a consistent truncation, as we have explained in \cite{ASW}.

For convenience, let us denote the five-dimensional scalars as $\Phi^i (x^I) = \{ p (x^I), x (x^I), g (x^I), \phi (x^I) a (x^I) b (x^I) \}$. Now, substituting the ansatz (\ref{10dmetric})-(\ref{F5}) into the 10d IIB action and integrating out the compact dimensions, one is left with an effective 5d action for the scalars $\Phi^i$ and the metric $g_{IJ}$ of the following form \cite{BHM}:
\be \label{S5d}
S = \int d^5 x \sqrt{- det g} \left[ - \frac{R}{4} + \frac{1}{2} \,G_{ij} (\Phi) \,\pd_{I} \Phi^i \pd^{I} \Phi^i + V (\Phi) \right] \, ,
\ee
where the only nonvanishing components of the sigma model metric $G_{ij} (\Phi)$ are:
\be \label{SigmaMM}
G_{pp} = 6 \,\,\,\,\, , \,\,\,\,\, G_{xx} = 1 \,\,\,\,\, , \,\,\,\,\, G_{gg} = \frac{1}{2} \,\,\,\,\, , \,\,\,\,\, G_{\phi \phi} = \frac{1}{4} \,\,\,\, ,
\ee
\be
G_{aa} = \frac{e^{-2g}}{2} \,\,\,\,\, , \,\,\,\,\, G_{bb} = \frac{P^2 e^{\phi - 2x}}{2}
\ee
and the potential $V (\Phi)$ is given by the expression:
\bea \label{Pot}
V (\Phi) &=& \frac{e^{-4p-4x}}{8} \!\left[ e^{2g} +(a^2-1)^2 e^{-2g} + 2a^2 \right] + \frac{a^2}{4} \,e^{-2g+8p} \nn \\
&+& P^2 \,\frac{e^{\phi-2x+8p}}{8} \!\left[ e^{2g} + e^{-2g} (a^2-2ab+1)^2 + 2(a-b)^2 \right] \nn \\
&-& \frac{e^{2p - 2x}}{2} \!\left[ e^g + (1+a^2) e^{-g} \right] + Q^2 \,\frac{e^{8p-4x}}{8} \,\,\, .
\eea

The field equations resulting from the above action are:
\bea \label{EoM}
\nabla^2 \Phi^i + {\cal G}^i{}_{jk} \,g^{IJ} (\pd_I \Phi^j) (\pd_J \Phi^k) - V^i &=& 0 \quad , \nn \\
- R_{IJ} + 2 \,G_{ij} \,(\pd_I \Phi^i) (\pd_J \Phi^j) + \frac{4}{3} \,g_{IJ} V &=& 0 \quad ,
\eea
where $\nabla^2 = \nabla_I \nabla^I$ and $V^i = G^{ij} V_j$ with $V_j \equiv \frac{\pd V}{\pd \Phi^j}$\,, while ${\cal G}^i{}_{jk}$ are the Christoffel symbols for the metric $G_{ij}$\,. Our goal is to find solutions of these equations of motion with a specific ansatz for the 5d metric.

\section{Metric ansatz and equations of motion}

We will be interested in solutions of (\ref{EoM}), for which the five-dimensional metric is of the form
\be
ds_5^2 = e^{2 A(r)} g_{\mu \nu} dx^{\mu} dx^{\nu} + dr^2 \,\, ,
\ee
where $\mu, \nu = 0,1,2,3$\,, \,and the fields $\Phi^i$ are functions of the radial direction only, i.e. $\Phi^i = \Phi^i (r)$. When $g_{\mu \nu} = \eta_{\mu \nu}$, this is the standard anstaz for the 5d metric and scalars of gravity duals of strongly coupled gauge theories living in {\it Minkowski} space. The different solutions (Klebanov-Strassler \cite{KS}, Maldacena-Nunez \cite{MN}, various recently found deformations of theirs \cite{NPP,ENP,EGNP}) differ by having different radial profiles for the scalars $\Phi^I (r)$ and different warpfactors $A(r)$. Also, there may be useful relations between subsets of scalars, that simplify the considerations. For instance, in the case of the Maldacena-Nunez solution one has \cite{BHM}:
\be
Q=0 \quad , \quad b=a \quad , \quad x=\frac{1}{2} g - 3p \quad , \quad \phi = -6p - g - 2 \ln P \,\, ,
\ee
leaving only three independent scalars $a(r)$, $g(r)$ and $p(r)$.

Here we are interested in a modification of the above ansatz, such that the 4d metric $g_{\mu \nu}$ is that of {\it de Sitter} space. Writing the $dS_4$ metric as
\be \label{dSm}
g_{\mu \nu} dx^{\mu} dx^{\nu} = -dt^2 + e^{2 \sqrt{\frac{\Lambda}{3}} \,t} d\vec{x}^2 \,\, ,
\ee
where $\Lambda > 0$ is the 4d cosmological constant, we find that the field equations (\ref{EoM}) acquire the form \cite{ASW}: 
\bea \label{EoMsyst}
(\Phi^i)'' + 4 A' (\Phi^i)' + {\cal G}^i{}_{jk} (\Phi^j)' (\Phi^k)' - G^{ij} V_j &=& 0 \nn \\
- \Lambda e^{-2A} + 4A'^2 + A'' + \frac{4}{3} V &=& 0 \nn \\
4 A'' + 4 A'^2 + 2 G_{ij} (\Phi^i)' (\Phi^j)' + \frac{4}{3} V &=& 0 \,\, ,
\eea
where we have denoted \,$'\equiv \pd_r$\,. One can immediatley see that the number of equations in (\ref{EoMsyst}) is greater by one than the number of unknown functions, $A(r)$ and $\Phi^i (r)$. So it might seem that it is not possible to solve the system of equations of motion with the desired metric ansatz. However, we showed in \cite{ASW} that as long as the warp factor is nontrivial, i.e. $A \neq const$, then one of the equations is dependent on the rest. To explain this dependence in more detail, let us introduce some useful notation. We denote by $E1$, $E2$ and $E3$ the three lines of equations in (\ref{EoMsyst}), respectively. Let us also denote by $dE$ the derivative of equation $E$ with respect to $r$. The dependence between the equations in the system (\ref{EoMsyst}) can then be described in the following way. Equation $E3-E2$ is linearly dependent on equations $E1$ and $d(4E2 - E3)$. Hence the system of independent equations to solve is actually: 
\bea \label{EoM2}
(\Phi^i)'' + 4 A' (\Phi^i)' + {\cal G}^i{}_{jk} (\Phi^j)' (\Phi^k)' - G^{ij} V_j &=& 0 \nn \\
-4 \Lambda e^{-2A} + 12 A'^2 + 4 V - 2 G_{ij} (\Phi^i)' (\Phi^j)' &=& 0 \,\, ,
\eea 
where the last line is $4E2-E3$.

Further simplification can be achieved by noticing that three of the six scalars $\Phi^i$ can be consistently set to zero. Namely, we showed in \cite{ASW} that setting
\be \label{gab}
g = 0 \quad , \quad a=0 \quad , \quad b = 0
\ee
solves automatically the $g$, $a$ and $b$ field equations. Using (\ref{gab}) and writing things out more explicitly, the system (\ref{EoM2}) becomes:
\bea \label{EoMspxphi}
p'' + 4 A' p' - V^p &=& 0 \,\,\, , \nn \\
x'' + 4 A' x' - V^x &=& 0 \,\,\, , \nn \\
\phi'' + 4 A' \phi' - V^{\phi} &=& 0 \,\,\, , \nn
\eea
\vspace{-0.58cm}
\be \label{EoMm}
4 \Lambda e^{- 2A} - 12 A'^2 + 12 p'^2 + 2 x'^2 + \frac{1}{2} \phi'^2 - 4 V \, = \, 0 \,\, ,
\ee
where
\be \label{Vs}
V = \frac{e^{-4p-4x}}{4} - e^{2p-2x} + \frac{e^{8p-2x+\phi}}{4} P^2 + \frac{e^{8p-4x}}{8} Q^2 \,\, .
\ee

\section{Solutions with $dS_4$ foliation}

We will find two kinds of solutions of (\ref{EoMm}). One of them will be obtained in a limit very similar to the one, in which the supersymmetric walking solutions of \cite{NPP} were found. The other class of solutions, basically, can be viewed as arising in a large $P$ limit. 

\subsection{Analytical solution}

The walking solutions of \cite{NPP} have an additional dynamical scale, compared to QCD-like theories. This is correlated with the fact that they have another constant parameter, denoted there by $c$, in addition to the number of colors $N_c$ of the dual gauge theory. The solutions of \cite{NPP} are actually obtained (at leading order) in the limit $c >\!\!> N_c$\,. Here we will explain that a similar limit exists for the system (\ref{EoMspxphi}) as well. 

To begin, let us rewrite the three scalars as:
\be
p (r) = \tilde{p} (r) + p_0 \quad , \quad x (r) = \tilde{x} (r) + x_0 \quad , \quad \phi (r) = \tilde{\phi} (r) + \phi_0 \,\, ,
\ee
where $p_0 = const$, $x_0 = const$ and $\phi_0 = const$. Then the potential (\ref{Vs}) acquires the form:
\be \label{VN}
V = \frac{1}{N_p^4 N_x^4} \,\frac{e^{-4\tilde{p}-4\tilde{x}}}{4} - \frac{N_p^2}{N_x^2} \,e^{2\tilde{p}-2\tilde{x}} + \frac{N_p^8 N_{\phi} P^2}{N_x^2} \,\frac{e^{8\tilde{p}-2\tilde{x}+\tilde{\phi}}}{4} + \frac{N_p^8 Q^2}{N_x^4} \,\frac{e^{8\tilde{p}-4\tilde{x}}}{8} \,\, .
\ee
where for convenience we have introduced the notation
\be
N_p \equiv e^{p_0} \quad , \quad N_x \equiv e^{x_0} \quad, \quad N_{\phi} \equiv e^{\phi_0} \,\, .
\ee
Now, recall that in ten dimensions $P$ and $Q$ are the coefficients of the $F_3$ and $F_5$ fluxes respectively. Taking $Q=0$ and noting that $P = \frac{N_c}{4}$, where $N_c$ is the number of D5 branes that are sources of $F_3$ flux, we have a system of the same kind as in \cite{NPP}. We will see below that this system has a simple analytic solution in a limit similar to the limit $c >\!\!> N_c$ relevant for the walking solution of \cite{NPP}. 

Indeed, let us consider the limit $\tilde{c} >\!\!> P$, where for convenience we have denoted $\tilde{c} \equiv N_p^{-3} N_{\phi}^{-1/2}$, together with $N_p^3 >\!\!> N_x^{-1}$. In this limit the potential reduces to 
\be \label{Vapprox}
V \approx - \,\frac{N_p^2}{N_x^2} \,e^{2\tilde{p}-2\tilde{x}} \,\, .
\ee
For simplicity of notation, let us write this as
\be \label{Vapp}
V \approx - \,N^2 \,e^{2p-2x} \,\, .
\ee
In other words, we denote $N^2 \equiv N_p^2 / N_x^2$ and, also, from now on we drop the tildes of the scalars. With this form  of the potential, one can easily see that taking 
\be \label{solphix}
\phi = 0 \qquad {\rm and} \qquad x = - 6p 
\ee
solves the $\phi$ field equation and transforms the equation for $x$ into the $p$ equation of motion \cite{ASW}. So we are left with a single independent scalar, namely $p(r)$. 

Now, upon substituting (\ref{solphix}) into (\ref{EoMm}), we find the following coupled system:
\bea \label{syst}
p'' + 4 A' p' + \frac{N^2}{3} e^{14p} &=& 0 \nn \\
4 \Lambda e^{- 2A} - 12 A'^2 + 84 p'^2 + 4 N^2 e^{14p} &=& 0 \,\, .
\eea
This system can be solved analytically by making the ansatz
\be \label{Apc0}
A(r) = - 7 p(r) + const \,\, .
\ee
For details see \cite{ASW}. The solution is given by
\bea \label{solpA}
p(r) &=& - \frac{1}{7} \,\ln (r+C) - \,\frac{1}{14} \,\ln \!\left( \frac{7 N^2}{9} \right) \,\, , \nn \\
A(r) &=& \ln (r+C) \,+ \,\frac{1}{2} \,\ln \!\left(\frac{7 \Lambda}{9}\right) \,\, ,
\eea
where $C$ is an integration constant.

It is worth noting that the functional form of the solution (\ref{solpA}) is consistent with ALD (asymptotically linear dilaton) behavior, as is the case also for \cite{NPP}. This was to be expected as they both belong to the same class of 10d systems, as explained in the beginning of this subsection. Since the ALD terminology may be a bit confusing, let us recall here, for the benefit of the reader, that in Einstein frame ALD behavior of a scalar field $\Phi (r)$ means $\Phi \sim c_1 \ln r + c_2 + ...$\,, where $c_{1,2} = const$ and the dots are terms subleading in powers of large $r$; see \cite{MMcN}. On the other hand, the name ``linear" arises from the behavior in string frame, which is $\Phi \sim r$, naturally.\footnote{See \cite{MV}, for example, for more details on ALD symptotics in Einstein vs string frames.} Beyond ALD behavior and the limit $\tilde{c} >\!\!> P$, there is yet another deep similarity between our solution and the walking backgrounds of \cite{NPP}. Namely, in both cases, at leading order, the scalars $g$, $a$ and $b$ vanish. All of that raises the natural question whether our solution could be understood as arising from a suitable nonsupersymmetric deformation of the supersymmetric walking backgrounds. And, more importantly, what that deformation would mean in the dual strongly coupled gauge theory. We leave investigating this interesting issue for the future.

\subsection{Numerical solutions}

Another class of solutions to (\ref{EoMm}) can be found in the $P >\!\!> 1$ limit with $Q=0$. In this approximation the potential is:
\be \label{VP}
V \approx \frac{e^{8p-2x+\phi}}{4} P^2 \,\, .
\ee
Equivalently, this leading form of the potential can be obtained from (\ref{VN}) by taking suitable relations between the various constants there, similarly to the limits $\tilde{c} >\!\!> P$ and $N_p^3 >\!\!> N_x^{-1}$ in the previous subsection. Regardless of the interpretation, we will now consider the case, in which the leading term in the potential is given by (\ref{VP}).

In that case, it is easy to verify that by setting
\be \label{xphip}
x = - \frac{1}{2} \phi \qquad {\rm and} \qquad \phi = 3p 
\ee
one can reduce the system of equations of motion to:
\bea \label{systPn}
p'' + 4 A' p' - \frac{1}{3} \,e^{14p} P^2 &=& 0 \,\,\, , \nn \\
4 \Lambda e^{-2 A} - 12 A'^2 + 21 p'^2 - e^{14p} P^2 &=& 0 \,\,\, .
\eea
See \cite{ASW} for more details on why the ansatz (\ref{xphip}) is consistent with the field equations and how it leads to (\ref{systPn}). Unfortunately, the system (\ref{systPn}) cannot be solved by making an ansatz similar to (\ref{Apc0}). So we have not been able to find an analytical solution. However, we succeeded in finding two classes of numerical solutions.  

To explain how to obtain those, let us introduce the following convenient notation:
\be 
H^2 \equiv \frac{\Lambda}{3} \,\,\, ,
\ee
so that the scale factor in (\ref{dSm}) becomes just $e^{Ht}$, as well as
\be \label{sub1}
A(r) = \ln \left( \frac{H}{B(r)} \right) \qquad {\rm and} \qquad p(r) = \frac{1}{7} \ln\left( \frac{q(r)}{P} \right) \,\, .
\ee
In terms of the $B(r)$ and $q(r)$ introduced above, equations (\ref{systPn}) become \cite{ASW}:
\bea\label{SystF3}
\frac{q''}{q^2\,q'}&=&\frac{q'}{q^3}+\frac{7\,q}{3\,q'}+\frac{4\,B'}{q^2 B} \,\, ,\nn\\
\frac{3\,q'{}^2}{7\,q^2}&=&\frac{12\,B'{}^2}{B^2}+q^2-12\,B^2 \,\, .
\eea
Now, one can solve the first equation in (\ref{SystF3}) algebraically for $B'/B$. Substituting this into the second equation, one finds and expression for $B$ entirely in terms of $q$, $q'$ and $q''$. Finally, substituting this expression for $B$ back into the first equation of (\ref{SystF3}), one ends up with a third order ODE for $q(r)$. This ODE has the form $F_1 \times F_2 = 0$, where each of the factors is an expression in terms of $q$ and its derivatives; see \cite{ASW}. One can show that the equation $F_1 = 0$ does not have a solution. Hence, the equation to solve, that one is left with, is $F_2 = 0$. Written out in more detail, this is \cite{ASW}: 
\be
q^{\prime \prime \prime}=\frac{49\,q^6}{36\,q'}+\frac{17}{2}\,q^2\,q'-\frac{9\,q'{}^3}{28\,q^2}-\frac{7\,q^3\,q''}{2\,q'}+\frac{q'\,q''}{2\,q}+\frac{5\,q''{}^2}{4\,q'} \,\,\, .
\ee

We were able to simplify this equation and solve it numerically by introducing a new function $T (\ln q)$ via the relation:
\be\label{qT}
q'(r)= \pm \frac{7}{3} \,q^2 \,\sqrt{1+T} \,\,\, .
\ee
For each of the $\pm$ signs in (\ref{qT}), we have found a separate class of numerical solutions. See \cite{ASW} for the technical details. 
\begin{figure}[htbp]
\begin{center}
\includegraphics[width=3.6in]{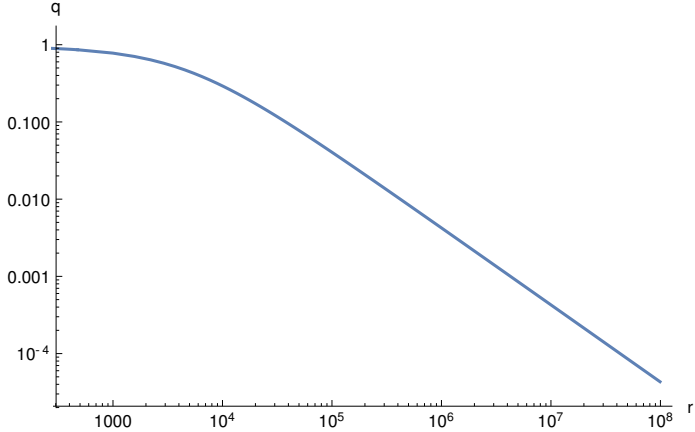}
\caption{The function $q(r)$ for integration constant $t = 0.1$.}
\end{center}
\end{figure}
In that work we illustrated the results graphically for the case of the plus sign. Here, instead, let us plot the numerical solutions for the minus sign in (\ref{qT}). In Figure 1 we show the graph of $q(r)$ for a certain value of an integration constant $t$, coming from the auxiliary function $T$. 
\begin{figure}[htbp]
\begin{center}
\includegraphics[width=3.6in]{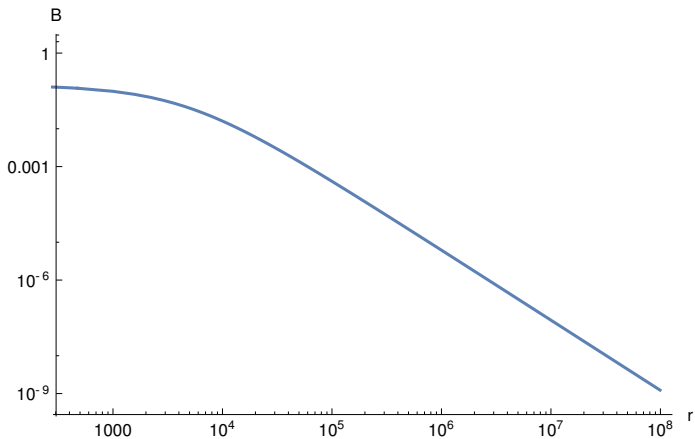}
\caption{The function $B(r)$ for integration constant $t = 0.1$.}
\end{center}
\end{figure}
In Figure 2 we plot $B(r)$ for the same value of the integration constant. These figures complement the analytical considerations in \cite{ASW}, supporting the same conclusion reached there. Namely, the functions $q(r)$ and $B(r)$ take only finite values for the entire $[0,\infty)$ range of the radial variable $r$ \cite{ASW}. Hence this solution is regular everywhere.

On the other hand, in the case of the plus sign in (\ref{qT}), the numerical solutions for $q$ and $B$ are such that the resulting functions $p(r)$ and $A(r)$ are finite at $r=0$, but logarithmically divergent at a certain $r=r_{max}$; see \cite{ASW}.

\section{Time-dependent deformations}

The solutions described in the previous section have a 5d metric of the form:
\be \label{MetricdS}
ds_5^2 = e^{2 A(r)} \left( -dt^2 + e^{2 H t} d\vec{x}^2 \right) + dr^2 \,\, ,
\ee
with $H = const$. They can provide a useful playground for studying strongly coupled gauge theories in de Sitter space in the vein of \cite{GIN}. More interestingly though, one can use these solutions as a starting point for building gravity duals of glueball inflation, as we will elaborate below. 

First, however, let us point out that glueball inflation, or more broadly inflation driven by a {\it composite} scalar in a strongly coupled gauge sector, was proposed already in \cite{CJS,BCJS}, within a purely field theoretic framework. The motivation was that such models would not suffer from the $\eta$-problem plaguing the standard ones, in which the inflaton is a fundamental scalar. Recall that the $\eta$-problem refers to the following. One of the inflationary slow roll parameters, denoted by $\eta$, is proportional to the inflaton mass. Therefore, if the inflaton is a fundamental scalar, the $\eta$ parameter receives quantum corrections, which spoil the slow roll approximation \cite{DB}. On the other hand, if the inflaton is a composite state in a strongly coupled gauge theory, then its mass is dynamically fixed and so there is no $\eta$-problem. For this reason, it is of great importance to develop calculable models of composite inflation and, in particular, of glueball inflation.

To obtain an inflationary spacetime, instead of the pure $dS_4$ space in the $(t,\vec{x})$ part of the metric (\ref{MetricdS}), we would like to have a solution with a time-dependent Hubble parameter $H$. This dependence, however, should be such that the following slow roll condition is satisfied \cite{DB}:
\be \label{HdotH21}
- \frac{\dot{H}}{H^2} <\!\!< 1 \,\, .
\ee
The expression on the left-hand side of this inequality, often denoted by $\varepsilon$, is the other slow roll parameter, in addition to $\eta$. The condition (\ref{HdotH21}) clearly means that $H (t)$ is a rather {\it slowly varying} function. Hence we can look for solutions of this kind by searching for small time-dependent deviations from an exact solution with a 4d de Sitter space. For such solutions to exist, the ansatz for the scalars also has to change to $\Phi^i = \Phi^i (t,r)$. We hope to report soon on progress in this directions. 

Finally, let us comment on why solutions of this kind would give models of glueball inflation. Recall that, in the context of the gauge/gravity duality, the scalars $\Phi^i$ represent glueballs in the dual gauge theory. Since a solution of the 5d consistent truncation of type IIB, which has an inflating 4d spacetime, would be supported by (some) nontrivial scalar fields $\Phi^i$, that means that the inflationary expansion in four dimensions is driven by a glueball. And, once one knows the function $H(t)$ for such a solution, one can immediately predict the inflationary observables $n_s$ and $r$, the scalar spectral index and the tensor-to-scalar ratio respectively, since the scalar and tensor power spectra are entirely determined by $H(t)$ and $\dot{H}(t)$; see \cite{DB}. Hence, indeed, such a solution would provide a model of glueball inflation.

\section*{Acknowledgments}

L.A. is grateful for partial support from the European COST Action MP-1210 and the Bulgarian NSF grant DFNI T02/6 during the completion of this work.

\end{document}